\documentclass[
aps,
pre,
twocolumn, 
showpacs, 
nopreprintnumbers, 
amsmath,amssymb,
letteraper
]{revtex4}

\usepackage{graphicx}
\usepackage{dcolumn}
\usepackage{bm}

\begin{document}

\preprint{WHU-PHY-0301}

\title{
Modified  conjugated gradient method for diagonalising large matrices
}


\author{Quanlin \surname{Jie}}
\email[E-mail: ]{qljie@whu.edu.cn}
\affiliation{%
Department of Physics, Wuhan University,
Wuhan 430072, P. R. China
}%

\author{Dunhuan \surname{Liu}}
\affiliation{%
Department of Physics, Wuhan University,
Wuhan 430072, P. R. China
}%


\date{\today}

\begin{abstract}
We present an iterative method to diagonalise large matrices. The basic
idea is the same as the conjugated gradient (CG) method, i.e, minimizing
the Rayleigh quotient via its gradient and avoiding reintroduce errors to
the directions of previous gradients.
Each iteration step is to find lowest eigenvector of the matrix in a subspace
spanned by the current trial vector and the corresponding gradient of the
Rayleigh quotient, as well as some previous trial vectors. The gradient,
together with the previous trail vectors, play a similar role of the conjugated
gradient of the original CG algorithm. Our numeric tests indicate that this
method converges significantly faster than the original CG method. And the
computational cost  of one iteration step is about the same as the original
CG method. It is suitably for first principle calculations.
%
\end{abstract}

\pacs{02.70.-c, 31.15.Ar, 31.15.-p, 95.75.Pq}
\keywords{Diagonalising large matrices, conjugated gradient method,
        computational algorithm}
\maketitle

\section{introduction}

In first principle calculations, such as band structure calculations, atomic
and molecular structure calculations, one of the basic tasks is to find several
lowest eigenvalues and the corresponding eigenvectors iteratively (and very
often self-consistently) of the effective Hamiltonian~\cite{1}. The matrix
dimension of the Hamiltonian may range from tens of thousands to several
millions, and one may need up to several thousands of the lowest eigenvectors
of the effective Hamiltonian. Diagonalising of matrices in such a scale needs
considerable CPU time and memories. It is one of the major numerical costs in
the first principle calculations, and the efficiency of the algorithm is
crucial for the performances of the whole program. There are many efforts to
improve the algorithm~\cite{2,3,4,5,6}.

Among widely used algorithms, such as Lanczos~\cite{5,7}, Dividson~\cite{8},
relaxation method~\cite{4,9},
DIIS ( Direct Inversion in the Iterative Subspace, which minimizes all matrix
elements between all trial vectors)~\cite{10}, and its later version
RMM-DIIS~\cite{2,11} (RMM stands for residual minimization, i.e.,
minimizing the norm of the residual vector in iterative subspace), the
conjugated gradient (CG) method~\cite{1,2} is a valuable tool to find a set of
lowest eigenvectors of a large matrix. Briefly speaking, to obtain the lowest
eigenvector of a matrix $H$ for the general form eigenvalue problem
\begin{equation}
H|\psi\rangle=E S |\psi\rangle,
\end{equation}
the CG method iteratively minimizes the Rayleigh quotient
\begin{equation}
E_n=\frac{\left\langle\phi_n\left|H\right|\phi_n\right\rangle}
{\langle\phi_n|S|\phi_n \rangle},
\end{equation}
where $S$ is the overlap matrix, and $|\phi_n\rangle$ is a refined trial
vector at step $n$. Each iteration step is to search the  minimization point in
the direction of the conjugated gradient which is a combination of the current
gradient and previous conjugated gradient.  One can obtain higher eigenvectors
in the same way, provided to keep the trial vector orthogonal to the lower
eigenvectors. In practical calculations, the CG method is stable and reasonably
efficient in many cases, and it is easy to implement. The iteration procedure
needs only to store the trial vector and its gradient, as well as one previous
conjugated gradient.

Conjugated gradient method is originally designed to minimize positive definite
quadratic functions iteratively. In $n$-th step of iteration, The CG method is
equivalent to find a minimum in a $n$-dimensional subspace spanned by
the initial trial vector and the subsequent $n-1$ gradients of the quadratic
function. Due to special properties of a quadratic function, one needs only do
the minimization in a two dimensional space spanned by current state and the
conjugated gradient which is a combination of current gradient and last step's
conjugated gradient. In principle, one needs at most $N$ steps to obtain final
solution in a $N$ dimensional space. Practical calculations usually needs
more steps
due to round off errors. The conjugated gradient method is virtually the most
effective method to minimize a quadratic function iteratively. And it is a
formally established algorithm to solve the linear algebraic equation.

For general functions, such as Rayleigh quotient, there are several ways to
define the conjugated gradient, and the behaviors of conjugated gradient
algorithm are unclear. However, near an exact minimum point, any
function behaves like a quadratic function. If one starts with a good guess,
one may find solution very quickly. This partially explains the successes of
the conjugated gradients method in diagonalising a large matrix.

\section{The modified conjugated algorithm}

Our method is based on the following two observations:

Firstly,
each iteration step of minimizing the Rayleigh quotient by CG algorithm is
equivalent to find lowest eigenvector in a two dimensional subspace. The
subspace at $n$-th step is spanned by the current state $|\phi_n \rangle$, and
the conjugated gradient $|F_n\rangle$. Note that the conjugated gradient
$|F_n\rangle$ is a combination of the gradient of $n$-th step's Rayleigh
quotient, and the $(n-1)$-th step's conjugated gradient $|F_{n-1}\rangle$. One
may expect a better result at $n$-th iterative step by finding lowest
eigenvector in a three dimensional subspace spanned by $|\phi_n\rangle$,
$|G_n\rangle$, and $|F_{n-1}\rangle$, where $|G_n\rangle$ is the gradient of
the Rayleigh quotient at $n$-th step.

Secondly, we
note that, within the CG algorithm, $|\phi_n\rangle$ is a combination of
$|\phi_{n-1}\rangle$ and $|F_{n-1}\rangle$. Thus the three dimensional subspace
spanned by $|\phi_n\rangle$, $|G_n\rangle$, and $|F_{n-1}\rangle$ is the same
as the subspace spanned by $|\phi_n\rangle$, $|G_n\rangle$, and
$|\phi_{n-1}\rangle$. This means that one may obtain a better result
$|\phi_{n+1}\rangle$ at $n$-th step by replace the $n$-step iteration of CG
algorithm with finding lowest eigenvector at the three dimensional subspace
spanned by $|\phi_n\rangle$, $|G_n\rangle$, and $|\phi_{n-1}\rangle$. Of
course, the result will further improved if one finds the lowest eigenvector in
a larger subspace spanned by $|\phi_n\rangle$, $|G_n\rangle$,
$|\phi_{n-1}\rangle$, $\cdots$, $|\phi_{n-m+2}\rangle$.

The above observations indicate that one may improve the efficiency of the CG
algorithm by replacing each iteration step of the CG algorithm with finding
lowest eigenvector in a small subspace spanned by the current vector
$|\phi_{n}\rangle$ and the corresponding gradient $|G_{n}\rangle$, as well as
some previous vectors $|\phi_{n-1}\rangle$, $|\phi_{n-2}\rangle$, $\cdots$. In
our numeric tests, the effect is significant in many cases. Since diagonalising
a small matrix of several dimension is very cheap numerically, each step's
numeric cost of the modified version is about the same as the original CG
algorithm.

Practical implementation of the modified conjugated gradient method is similar
to that of original CG method. For finding one single lowest eigenvalue and its
corresponding eigenvector, it goes through the following steps:
\begin{enumerate}

\item

Choose the dimension $M$ of the iteration subspace, and the maximum iteration
step $N_{max}$. In our numerical
test, it is enough to set the dimension $M\le 10$. In many case, $M=3$ works
quite well. In this case, the 3 dimensional subspace is spanned by current
trial vector $|\phi_n\rangle$, the corresponding gradient $|G_n\rangle$ and
one previous trial vector $|\phi_{n-1}\rangle$ obtained in the last step.

\item

Choose an initial normalized trial vector $|\phi_0\rangle$,
$\langle\phi_0|S|\phi_0\rangle=1$; and calculate the expectation value
(Rayleigh quotient) $E_0=\langle\phi_0|H|\phi_0\rangle$.

\item

For $n=0,\, 1,\, 2,\, \cdots$, $N_{max}$, do the following iteration loop to
refine the trial
vector from $|\phi_0\rangle$ to $|\phi_1\rangle$, $|\phi_2\rangle$, $\cdots$:
\begin{enumerate}

\item

Calculate the gradient of the Rayleigh quotient
\begin{equation}
|G_n\rangle = H|\phi_n\rangle - E_n|\phi_n\rangle.
\end{equation}
Here the refined trial function $|\phi_n\rangle$ is normalized at the end of
each iteration.

\item

In the $m$ dimensional subspace spanned by $|\psi_1\rangle = |G_n\rangle$,
$|\psi_2\rangle = |\phi_{n}\rangle$, $|\psi_3\rangle = |\phi_{n-1}\rangle$,
$\cdots$, $|\psi_m\rangle = |\phi_{n-m+2}\rangle$, calculate the matrix
elements of the matrix $H$, $\mathcal{H}_{ij}=\langle\psi_i|H|\psi_j\rangle$,
and the overlap matrix of the basis vector
$\mathcal{S}_{ij}=\langle\psi_i|S|\psi_j\rangle$. Here, the dimension $m=n+1$
if $n+1\leq M$, otherwise $m=M$, i.e., in the first $M-1$ loops, the subspace
has only $n+1$ basis vector.

\item

find the lowest eigenvalue $\epsilon$ and eigenvector $\varphi$ for the general
form eigenvalue problem
\begin{equation}\label{3}
\mathcal{H}\varphi=\epsilon \mathcal{S}\varphi.
\end{equation}

\item

From the above eigenvector $\varphi$, construct the refined trial vector
$|\phi_{n+1}\rangle$,
\begin{equation}\label{4}
|\phi_{n+1}\rangle=\sum_{i=1}^m\varphi_i |\psi_i\rangle,
\end{equation}
and calculate the expectation value
$E_{n+1}=\langle\phi_{n+1}|H|\phi_{n+1}\rangle$.

\item

If $|E_{n+1}-E_{n}|$ is less than a required value or $n>N_{max}$, stop the
iteration loop, otherwise continue the iteration loop.

\end{enumerate}
\end{enumerate}

Impose a maximum iteration
step is necessary in many cases. For example, in self-consistent calculations,
one needs to update the Hamiltonian after some steps of iterations.
The trial vector can be chosen, in principle, arbitrarily, provided
it is not orthogonal with the lowest eigenvector. However,
even if the initial trial vector does accidently orthogonal to the lowest
eigenvector, due to the numeric round off errors in the iterations,
one can always arrive the lowest eigenvector.

Check the convergence is usually testing the difference between the trial
vector and its refined version after an iteration. In our numeric tests, check
the difference between two consecutive trial vectors' Rayleigh quotients
also works well. And it is numerically faster.


For large matrices, calculation of the gradient is a main numeric task in
each loop of iteration. It involves a multiplication of matrix and vector.
Other numeric costs are mainly the calculation of the matrix elements
$\mathcal{H}_{ij}$ and $\mathcal{S}_{ij}$ in the small subspace , as well as
the combination of the gradient and previous trial vectors to form a refined
trial vector. The numeric cost of diagonalising the small matrix
$\mathcal{H}$ is almost
nothing as compared with other operations. In each loop of iteration, the
subspace changes two basis vectors, i.e., the current gradient $|G_n\rangle$
replaces the previous one $|G_{n-1}\rangle$, and the refined trial vector
$|\phi_n\rangle$ replace the old one $|\phi_{n-m+2}\rangle$. One needs only to
calculate the matrices elements $\mathcal{H}_{ij}$ and $\mathcal{S}_{ij}$
related to the two vectors in each iteration loop. If the subspace is three
dimensional, the numerical cost of one iteration loop is about the same as that
of original CG method.


After finding the lowest eigenvector, one can  find the second lowest one in a
similar way. One starts with a trial vector orthogonal to the lowest
eigenvector, and in following iterations, gradients of the Rayleigh quotient,
as well as the updated trial vectors, must be kept orthogonal to the lowest
eigenvector. Similarly, after working out $k$ lowest eigenvectors, the $k+1$
eigenvector can be worked out by maintaining the orthogonality with $k$ lower
eigenvectors.

In this strict sequential procedure, the accuracy of lower eigenvectors affect
the higher ones. A remedy to this problem, according to Ref.~\cite{2}, is
re-diagonalising the matrix in the subspace spanned by the refined trial
vectors, which is referred as subspace rotation in~\cite{2}. After this
subspace rotation, one can use these resulted vector as trial vectors for
further iteration to improve the accuracy. In practical implementations, we
only iterate every trial vector for some steps, then perform a subspace
rotation. The convergence check is to test the eigenvalues differences between
two consecutive subspace rotation.  This procedure
improves the overall efficiency. In
Ref.~\cite{2}, there is a detailed discussion on the role of the subspace
rotation.

\section{Numerical results}

We test the efficiency of the above outlined algorithm by comparing
its performance with other algorithms for various matrices.
In all cases, the modified CG algorithm outperforms the original CG algorithm.
We observe significant improvement to the convergence rate in many cases.

As an illustration, we show in figure 1 a typical result for a banded matrix
with bandwidth $2L$. The matrix's diagonal element is  $a_{ii}=2\sqrt{i}-a$,
and its off-diagonal elements within the band width is a constant $a_{ij}=a$.
Due to its simple form and its relation with Hamiltonian describing the pairing
effects, this matrix has been investigated by some other authors, see,
e.g.~\cite{4}. Here we choose the matrix's dimension $N=200000$ with
half-bandwidth $L=300$, the parameter $a$ is set to be 20. For finding first 8
lowest eigenvectors, the modified CG algorithm converges within 100 steps with
an accuracy of machine's precision limit. It is more than three times faster
than the original CG algorithm. As a comparison, we also show the result for
the block Lanczos method~\cite{7}, as well as the steepest decent method. In
Figure 1, the convergence rate of one iteration step is defined as the relative
error of the two consecutive Rayleigh quotients,
$(E_{n}-E_{n-1})/[(E_{n}+E_{n-1})/2]$, where $E_{n-1}$ and $E_n$ are two
consecutive Rayleigh quotients. When every eigenvalue reaches the required
accuracy, we perform a subspace rotation and repeat the iteration. Convergence
is to test the corresponding relative error for every eigenvalue between two
consecutive rotations. In our implementation, the maximum iteration number
$N_{max}=500$, i.e., we go at most 500 steps of iteration for each trial vector
before performing a subspace rotation.

\begin{figure}[h]
\includegraphics[angle=0,width=7.6cm,clip]{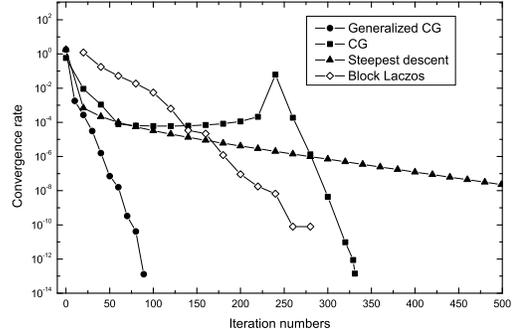}
\label{fig1}
\caption{
Convergence rate of the modified conjugated gradient method in comparing
with other algorithms.
}
\end{figure}

In the above calculations, we use a 3-dimensional iteration subspace for the
modified CG algorithm, i.e., the subspace is constitute of the current trial
vector, its corresponding gradient, as well as one previous trial vector. In
such case, each iteration step needs to calculate one gradient, and some
combinations of the three vectors, as well as solving a 3-dimensional
eigenvalue problem. From the above argument, when the iteration subspace is
3-dimensional, the numeric cost of each iteration step is almost the same as
that of the original CG method. For the block Lanczos algorithm, however, to
ensure a reasonable convergence rate, the iteration subspace is 50 dimensional,
i.e., one needs to calculate 50 gradients for each iteration step. To our
experience, on step of Lanczos iteration needs longer CPU time than 50 steps of
the modified CG method. Thus, one Lanczos step is counted as 50 steps in Figure
1.


In the 3-dimensional iteration subspace spanned by $\{|G_n\rangle$,
$|\phi_n\rangle$, $|\phi_{n-1}\rangle\}$ the gradient vector $|G_n\rangle$,
together with the previous trial vector $|\phi_{n-1}\rangle$, play the same
role as that of the conjugate gradient in the minimization of a quadratic
function. This is especially the case when the Rayleigh quotient closes to the
minimum point, i.e., it is approximately a quadratic function of the iteration
trial vector. In fact, without the previous trial vector $|\phi_{n-1}\rangle$,
the lowest eigenvector obtained in the 2-dimensional subspace spanned
by$\{|G_n\rangle,\ |\phi_n\rangle\}$ is just the result of steepest descent
method. By
including one previous trial vector which contains information about previous
gradients, one is able to prevent reintroduction of errors to the refined trial
vector in the direction of previous gradients. This is the reason we call this
method as modified CG algorithm.


On the other hand, in the sense of relaxation algorithm for finding lowest
eigenvector~\cite{4,9}, the refined trial vector $|\phi_{n}\rangle$ at step
$n$, is an approximation to the lowest eigenvector of the matrix in the
subspace spanned by
$\{|\phi_{0}\rangle,\,\,|G_{1}\rangle,\,\,|G_{2}\rangle,\,\cdots,\,\,
|G_{n}\rangle\}$, which is equivalent to the subspace spanned by
$\{|\phi_{0}\rangle,\,\,|\phi_{1}\rangle,\,\,|\phi_{2}\rangle,\,\,\cdots,
\,|\phi_{n-1}\rangle,\,\,|\phi_{n}\rangle\}$. According to the relaxation
algorithm, to find the lowest eigenvector in the subspace spanned by the above
basis vectors, one starts from an initial trial vector $|\psi_{0}\rangle$, and
minimizes the Rayleigh quotient iteratively. Each iterative step is to
minimize the Rayleigh quotient in a two dimensional subspace spanned by the
(updated) trial vector, and one basis vector. The basis vector can be chosen
consecutively from the first one to the last one. After going through all basis
vectors, one continues the next round of iteration by choosing the first basis
vector as next basis vector. This iteration will converge after goes through
all basis vectors several rounds. Note that, if one starts with the first basis
vector $|\phi_{0}\rangle$ as initial trial vector, in the two dimensional
subspace spanned by two consecutive basis vector $|\phi_{i}\rangle$, and
$|\phi_{i+1}\rangle$, the second basis vector $|\phi_{i+1}\rangle$minimizes the
Rayleigh quotient. After going through all basis vector for one round, the
refined trial vector is $|\phi_{n}\rangle$, which represents an approximate
lowest eigenvector in the above subspace.

The above two factors explain the rapid convergence of the modified CG
algorithm. One consequence from the above arguments is that, if we increase the
dimension of the iteration subspace by including more previous trial vectors,
the convergence rate will not increase too much. In other words, one needs only
do the modified CG algorithm in a small iteration space. To our experience, one
needs at most 5-dimensional iteration subspace. In most cases, it is enough to
do the iteration in the 3-dimensional iteration subspace. Figure 2 shows our
numeric result to confirm this property of the modified CG algorithm. Here we
do the same calculation using different iteration subspace. The filled circle
connected line is the same as figure 1 with 3-dimensional iteration subspace,
and the filled square and triangle are results for 6 dimensional and 12
dimensional iteration subspace respectively. There is almost no difference
within 50 steps where the the convergence rate is about $10^{-8}$. One needs
almost the same iteration steps to arrive the final precision. However, the
3-dimensional iteration runs faster for each iteration step since it involves
less combination and production of the basis vectors that span the iteration
subspace.

\begin{figure}[h]
\includegraphics[angle=0,width=7.6cm,clip]{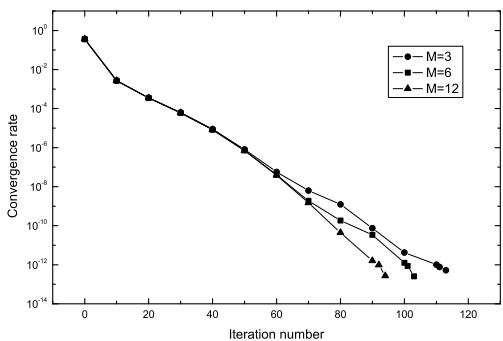}
\label{fig2}
\caption{
Convergence rates of the modified CG algorithm
for different dimensions of the iteration subspace.
}
\end{figure}


For some matrices or some properly chosen initial trial vectors, the Rayleigh
quotient are approximately quadratic functions of the trial vectors. In such
cases,  the modified CG algorithm converges in almost the same rate as the
original CG algorithm. And a trial vector $\phi_{n}$ at step $n$, is an almost
exact minimum in the subspace spanned by
$\{|\phi_{0}\rangle,\,|\phi_{1}\rangle,\,\cdots,\,|\phi_{n}\rangle,\,
|G_{n}\rangle\}$. We have encountered such cases in our numeric tests. In fact,
near a minimum, any function behaves like a quadratic function. Some matrices
with special structures also make the Rayleigh quotient like a quadratic
function in a quite large region of the vector space. For such matrices, the CG
method is indeed a very efficient method. Of cause, in any cases the modified
CG method always outperforms the original CG method.

The refined trial vector $|\phi_{n}\rangle$ becomes closer and closer to the
previous step's trial vector $|\phi_{n-1}\rangle$ when iteration closes to
final solution. In higher dimensional iteration, one may encounter (numerical)
degeneracy of basis vectors that span the iteration subspace. This problem is
easy to solve. One simple solution is to replace this step by an steepest
descent's step. Other more sophisticated way is to choose some independent
vectors from the basis vectors and do this step in a small subspace. Both
methods are easy to implement. In fact, one can detect the degeneracy when
solving the general form eigenvalue problem (\ref{3}) which can be conveniently
solved by the conventional Choleski-Householder procedure~\cite{12}.  If there
is a degeneracy, the Choleski decomposition of the overlap matrix $\mathcal{S}$
returns an error code. When this happens, one can simply redo this step with a
steepest descent step. Alternatively, one can use a more sophisticated Choleski
decomposition program that automatically chooses independent basis vector. In
doing so, one must adjust the the matrix element of $\mathcal{H}$
simultaneously. This two methods need almost the same numerical cost. Of
course, the first method is easy to implement.  In our numerical tests, there
is almost no degeneracy in the 3-dimensional iteration subspace.


It is straightforward to implement preconditioning treatment
for the modified CG algorithm. Preconditioning treatment can significantly
improve the convergence rate for matrices with large difference between
lowest and highest eigenvalues. Due to the fact that there is no need to
construct explicitly the conjugated gradient in the modified CG algorithm,
it is easier to implement the preconditioning treatment by direct
modifying each step's gradient. Since preconditioning treatment depends on
specific system, we don't go into more details about such topic.


The modified CG algorithm shares a common feature with many other iterative
methods of diagonalising matrices, such as Lanczos, Dividson, RMM-DIIS, and
relaxation method. In all these algorithms, one refines the trial vector in
iterative subspaces. What makes the modified CG algorithm different from other
algorithms is that the iteration subspaces are spanned by the trial vectors of
previous iteration steps, as well as the latest trial vector and its gradient.
The trial vectors of previous steps are already prepared, one needs only
calculating one gradient vector (and possibly does some preconditioning
treatment) to construct the basis vectors of the iterative subspace. Only two
basis vectors of the iterative subspace are different from previous one, it
needs only update two columns of the matrix elements in the iteration
subspace. By including previous trial vectors into the iterative subspace, one
avoids reintroduces errors to the trial vectors in the previous directions of
gradients. These properties of the iterative subspace make the modified CG
algorithm numeric efficient. And the common feature of the algorithm makes it
easy to implement.



It is easy to formulate block algorithm for the modified CG algorithm to find
several lowest eigenvectors simultaneously. For this end, one refines several
trial vectors at each iteration step. Here the iteration subspace includes all
current trial vectors, their gradients, and all trial vectors of some previous
steps. In this implementation, one needs to find several eigenvectors by
solving the general form eigenvalue problem (\ref{3}). Trial vectors obtained
in this way are automatically orthogonal with each other, and one needs no
additional subspace rotation.

However, one step of block algorithm usually needs more floating point
operations than sequentially processing each trial vector and maintaining
orthogonality between trial vectors by Schmidt orthogonalization method. This
is mainly because the block algorithm needs more flops to form the matrix
elements of $\mathcal{H}$ and the corresponding overlap matrix $\mathcal{S}$.
If one needs $n_0$ lowest eigen-solutions for $N$ dimensional matrix, the
block algorithm's iterative subspace is $M=m n_0$ dimensional with
$m=3,4,\cdots$.  Each step of block algorithm needs the following floating
point operations: (a) $n_0 NL$ flops for $n_0$ matrix multiplying vector
operations to obtain $n_0$ gradients, where $L\le N$ is the band width of the
matrix; (b) $2(m n_0)^2 N$ flops for the formation of the matrix elements of
$\mathcal{H}$ in the iterative subspace and the corresponding overlap matrix
$\mathcal{S}$; (c) An $O\left((m n_0)^3\right)$ floating point operations for
solving the general form eigen-value problem (\ref{3}); (d) $2mn_0^2 N$ flops
for combination the $mn_0$ basis vectors to form $n_0$ refined trial vectors.
Here, the flops in step (c) is negligible when $n_0<<N$.  The total flops of
one step block algorithm is $\sigma(m,n_0,N)=n_0 NL+2(m n_0)^2 N+2mn_0^2 N$.
If $n_0=1$, the above floating point operations $\sigma(m,1,N)=NL+2m^2 N+2m N$
is the flops for processing one trial vector in sequential algorithm. One the
other hand, sequentially processing each trial vector one round needs
$n_0\sigma(m,1,N)+4n_0^2 N$ flops. Here the second term is the flops to
maintain the orthogonality of trial vectors, including making gradients
orthogonal to previous trial vectors. Even including subspace rotation which
is performed after some rounds of sequential steps, the sequential
implementation needs less floating point operations than the block algorithm.

If $n_0$ is small, e.g., $n_0<10$, the difference of flops between block and
sequential algorithm is small. The block algorithm may be one choice in such
cases.  Like the block Lanczoz~\cite{7}, and block Dividson~\cite{8}, there
are some other ways to form the iterative subspace to implement the block
version of modified CG algorithm. For example, the iterative subspace may
contain only one gradient, plus all the current trial vectors and some
previous trial vectors. The choice of iterative subspace affects the
convergence properties which needs further investigations.  For large $n_0$,
e.g., $n_0>100$, to our experiences, block algorithm need more numeric cost
and is less efficient as compared with the above sequential implementation.
The dimension of the iteration subspace grows quickly with the number of
needed eigenvectors, and one needs more memory to store the basis vectors and
much more CPU time to solve the general from eigenvalue problem (\ref{3})
which increases drastically with the dimension of the iterative subspace.
Since lowest eigenvector usually converges faster that higher ones, the number
of iteration steps in a block algorithm is determined by the the vector with
slowest convergence rate.

\section{conclusions}

In summary, in the sense of conjugated gradient algorithm, we formulate an
iterative method to find a set of lowest eigenvalues and eigenvectors of a
matrix. This method minimizes the Rayleigh quotient of a trial vector via the
gradient of the Rayleigh quotient, and at the same time, prevents reintroduce
errors in the direction of previous gradients. We realize such idea by refining
the trial vectors in a special kind of iteration subspaces. Each iteration
subspace is spanned by the latest trial vector and the gradient of its Rayleigh
quotient, as well as some trial vectors of previous steps. Each iteration step
is to find lowest eigenvector in the iteration subspace. The gradient, together
with the previous trial vector, play the role of the conventional conjugated
gradient. In our numerical test, it is usual enough to include only one
previous trial vector, i.e., one needs only refining the trail vector in a
3-dimensional subspace. As compared to the conventional conjugated gradient
algorithm, which is designed to minimizes a general function, the current
method exploits special properties of eigenvalue problems, and thus converges
much faster in many cases. During iterations, the trial vector at the step $n$,
is an approximately lowest eigenvector in the subspace spanned by the initial
trial vector and $n$ subsequent gradient vectors. This is the reason of rapid
convergence rate. The easy implementation of this algorithm makes it suitable
for first principle calculations.

\bigskip

This work is supported in part by the National Natural Science Foundation,
the Research Fund of the State Education Ministry of China, and the Research
Fund of the Wuhan University. We thanks helpful discussions with Prof. W.
Wang.


\end{document}